\documentclass[10pt, conference]{IEEEtran}
\IEEEoverridecommandlockouts
\usepackage{cite}
\usepackage{amsmath,amssymb,amsfonts}
\usepackage{algorithmic}
\usepackage{graphicx}
\usepackage{textcomp}
\usepackage{xcolor}
\usepackage[bookmarks=false]{hyperref}
\usepackage{csquotes}
\usepackage[nolist,nohyperlinks]{acronym}
\usepackage{listings}
\ifCLASSOPTIONcompsoc
    \usepackage[caption=false,font=normalsize,labelfont=sf,textfont=sf]{subfig}
\else
    \usepackage[caption=false,font=footnotesize]{subfig}
\fi
\usepackage{xspace}

\newcommand{\Eg}{E.g.\xspace}
\newcommand{\eg}{e.g.\xspace}
\newcommand{\Ie}{I.e.\xspace}

\newcommand{\etal}{et al.\xspace}

\IEEEtriggeratref{38}

\makeatletter
\newcommand\autorefs[1]{\@first@ref#1,@}
\def\@throw@dot#1.#2@{#1}% discard everything after the dot
\def\@set@refname#1{%    % set \@refname to autoefname+s using \getrefbykeydefault
    \edef\@tmp{\getrefbykeydefault{#1}{anchor}{}}%
    \xdef\@tmp{\expandafter\@throw@dot\@tmp.@}%
    \ltx@IfUndefined{\@tmp autorefnameplural}%
         {\def\@refname{\@nameuse{\@tmp autorefname}s}}%
         {\def\@refname{\@nameuse{\@tmp autorefnameplural}}}%
}
\def\@first@ref#1,#2{%
  \ifx#2@\autoref{#1}\let\@nextref\@gobble% only one ref, revert to normal \autoref
  \else%
    \@set@refname{#1}%  set \@refname to autoref name
    \@refname~\ref{#1}% add autoefname and first reference
    \let\@nextref\@next@ref% push processing to \@next@ref
  \fi%
  \@nextref#2%
}
\def\@next@ref#1,#2{%
   \ifx#2@ and~\ref{#1}\let\@nextref\@gobble% at end: print and+\ref and stop
   \else, \ref{#1}% print  ,+\ref and continue
   \fi%
   \@nextref#2%
}
\makeatother

\begin{acronym}[WYSIWYG]
    \acro{a2m}[A2M]{architecture meta model}
    \acro{a3m}[A3M]{architecture meta meta model}
    \acro{adl}[ADL]{Architecture Description Language}
    \acro{api}[API]{application programming interface}
    \acro{cd}[CD]{continuous deployment}
    \acro{ci}[CI]{continuous integration}
    \acro{cli}[CLI]{Command Line Interface}
    \acro{di}[DI]{Dependency Injection}
    \acro{dsl}[DSL]{Domain Specific Language}
    \acro{ebnf}[EBNF]{extended Backus--Naur form}
    \acro{elk}[ELK]{Eclipse Layout Kernel}
    \acro{emf}[EMF]{Eclipse Modeling Framework}
    \acro{glsp}[GLSP]{Graphical Language Server Platform}
    \acro{id}[ID]{identifier}
    \acro{ide}[IDE]{Integrated Development Environment}
    \acro{json}[JSON]{JavaScript Object Notation}
    \acro{kdm}[KDM]{Knowledge Discovery Meta-Model}
    \acro{klighd}[KLighD]{KIELER Lightweight Diagrams}
    \acro{kieler}[KIELER]{Kiel Integrated Environment for Layout Eclipse Rich Client}
    \acro{loc}[LoC]{lines of code}
    \acro{lsp}[LSP]{Language Server Protocol}
    \acro{mde}[MDE]{Model Driven Engineering}
    \acro{mil}[MIL]{Module Interconnection Language}
    \acro{mof}[MOF]{Meta Object Facility}
    \acro{mvc}[MVC]{Model-View-Controller}
    \acro{osgi}[OSGi]{Open Services Gateway initiative}
    \acro{pm}[PM]{project model}
    \acro{spviz}[SPViz]{Software Project Visualization}
    \acro{ui}[UI]{user interface}
    \acro{uml}[UML]{Unified Modeling Language}
    \acro{vcm}[VCM]{view context model}
    \acro{vc2m}[VC2M]{view context meta model}
    \acro{vc3m}[VC3M]{view context meta meta model}
    \acro{vscode}[VS~Code]{Visual Studio Code}
    \acro{xmi}[XMI]{\acs{xml} Metadata Interchange}
    \acro{xml}[XML]{Extensible Markup Language}
\end{acronym}

\acused{api}
\acused{cd}
\acused{ci}
\acused{cli}
\acused{id}
\acused{ide}
\acused{json}
\acused{osgi}
\acused{ui}
\acused{uml}
\acused{vscode}
\acused{xml}

\lstdefinestyle{spvizmodel} {
morekeywords={package, SPVizModel, contains, connects},
otherkeywords={},
morecomment=[l]{//}, % l is for line comment
morecomment=[s]{/*}{*/} % s is for start and end delimiter
}

\lstdefinestyle{spviz} {
morekeywords={package, import, SPViz, shows, with, from, >, show, in, connect, via, and, connected},
otherkeywords={},
morecomment=[l]{//}, % l is for line comment
morecomment=[s]{/*}{*/} % s is for start and end delimiter
}

\lstdefinestyle{EBNF} {
morekeywords={},
otherkeywords={},
morecomment=[s]{?}{?} % s is for start and end delimiter
}

\begin{document}

\title{SPViz: A DSL-Driven Approach \\for Software Project Visualization Tooling %
    \thanks{This work has been supported by the project \emph{Visible Code},
        a cooperation between Kiel University and Scheidt \& Bachmann System Technik GmbH.}}

\author{\IEEEauthorblockN{Niklas Rentz and Reinhard von Hanxleden}
    \IEEEauthorblockA{Department of Computer Science\\
        Kiel University, Kiel, Germany\\
        \{nre,rvh\}@informatik.uni-kiel.de}
}

\maketitle

%%%%%%%%%%%%%%%%%%%%%%%%%%%%%%%%%%%%%%%%%%%%%%%%%%%%%%%%%%%%%%%%%%%%%%%%%%%%%%%%%%%%%%%%%%%%%%%%%%
% ------------------------------------------ Abstract ------------------------------------------ %
%%%%%%%%%%%%%%%%%%%%%%%%%%%%%%%%%%%%%%%%%%%%%%%%%%%%%%%%%%%%%%%%%%%%%%%%%%%%%%%%%%%%%%%%%%%%%%%%%%

\begin{abstract}
    For most service architectures, such as \ac{osgi} and Spring, architecture-specific tools allow software developers and architects to visualize otherwise obscure configurations hidden in the project files.
    Such visualization tools are often used for documentation purposes and help to better understand programs than with source code alone.
    However, such tools often do not address project-specific peculiarities or do not exist at all for less common architectures, requiring developers to use different visualization and analysis tools within the same architecture.
    Furthermore, many generic modeling tools and architecture visualization tools require their users to create and maintain models manually.

    We here propose a \acs{dsl}-driven approach that allows software architects to define and adapt their own project visualization tool.
    The approach, which we refer to as \ac{spviz}, uses two \acsp{dsl}, one to describe architectural elements and their relationships, and one to describe how these should be visualized.
    We demonstrate how \ac{spviz} can then automatically synthesize a customized, project-specific visualization tool
    that can adapt to changes in the underlying project automatically.

    We implemented our approach in an open-source library, also termed \ac{spviz}
    and discuss and analyze four different tools that follow this concept, including open-source projects and projects from an industrial partner in the railway domain.
\end{abstract}

%%%%%%%%%%%%%%%%%%%%%%%%%%%%%%%%%%%%%%%%%%%%%%%%%%%%%%%%%%%%%%%%%%%%%%%%%%%%%%%%%%%%%%%%%%%%%%%%%%
% ---------------------------------------- Introduction ---------------------------------------- %
%%%%%%%%%%%%%%%%%%%%%%%%%%%%%%%%%%%%%%%%%%%%%%%%%%%%%%%%%%%%%%%%%%%%%%%%%%%%%%%%%%%%%%%%%%%%%%%%%%
\section{Introduction}
\label{sec:introduction}
This is joint work with the industrial partner \emph{Scheidt \& Bachmann System Technik GmbH}
who has to maintain large software projects for a long duration,
a setting that is quite common in industry.
To maintain a good understanding of complex software architectures is a non-trivial task.
Furthermore, one regularly has to make new team members familiar with the existing software.
\emph{Diagrams} can aid understanding concrete connections and ideas
and the broader architecture of a system~\cite{ShahinLAB14, EadesZ96}.
However, it is still common practice to create such diagrams manually.
This requires significant maintenance effort~\cite{LientzST78}
and bears the risk of becoming inconsistent with the actual project.

\begin{figure}
    \centering
    \includegraphics[width=0.88\columnwidth]{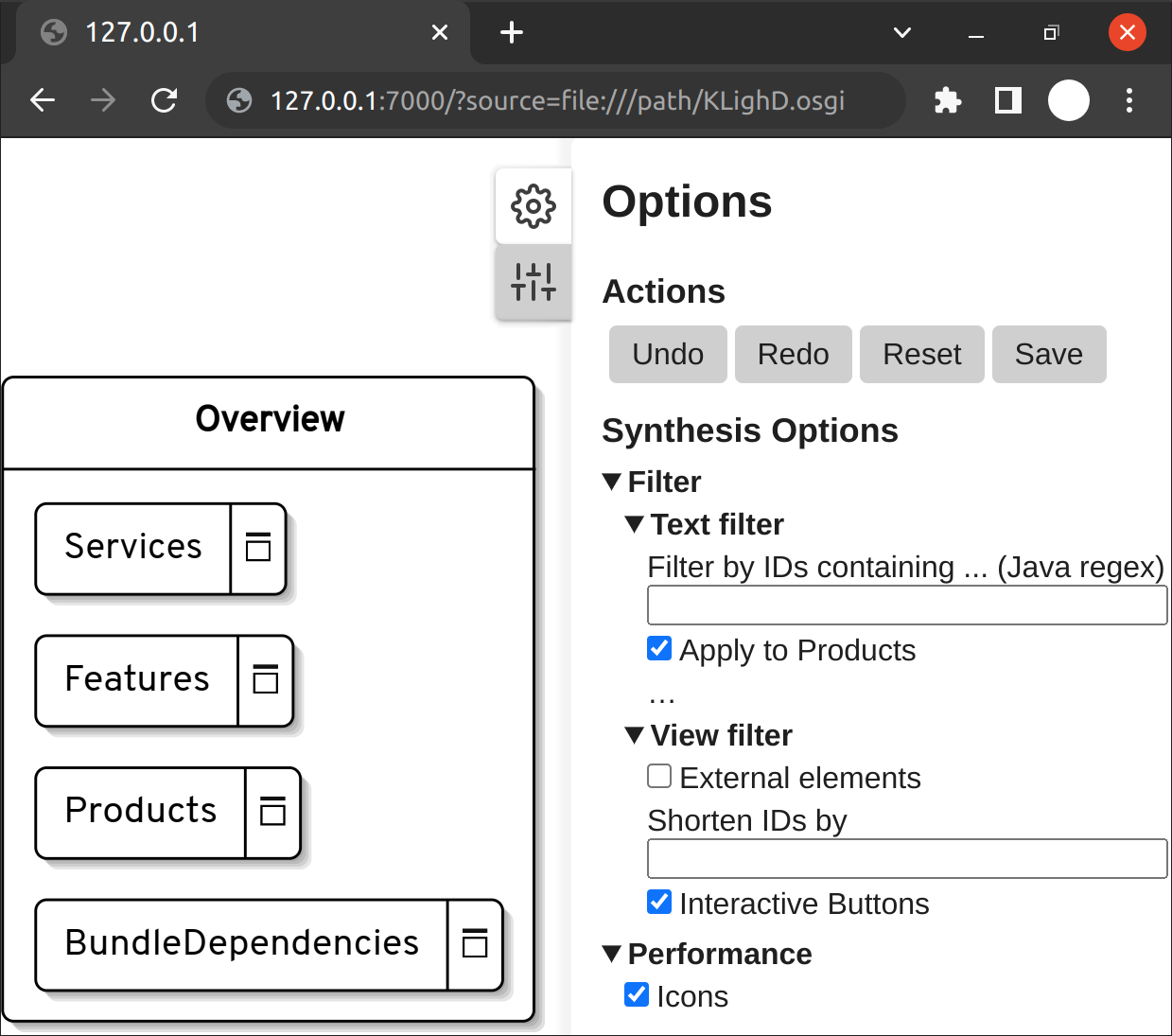}
    \caption{Screenshot of an architecture visualization tool synthesized by \ac{spviz}, here for \ac{osgi} projects. The overviews can be interactively expanded to show connections as shown in Figures \ref{fig:bundle-dep-view} and \ref{fig:category-connection-view}. The view can be customized with filters and interactive features in the open sidebar.}
    \label{fig:view-initial-web}
\end{figure}

There are multiple approaches to combat this issue.
There are tools to reverse-engineer the actual architecture of projects to compare it to the modeled architecture and adapt them to each other~\cite{MurphyNS01, TranGLH00}.
Other solutions require a direct inclusion of the architecture design into the actual source artifacts.
Such architecture inclusion is usually done via extended languages~\cite{AldrichCN02},
\acp{adl} that are developed alongside the architecture~\cite{BuchgeherW08},
or language-specific architecture systems such as \ac{osgi}\footnote{
    \ac{osgi}\texttrademark{} is a trademark of the \ac{osgi} Alliance in the US and other countries.
}
~\cite{Osgi20,RentzDvH20}.
For example, \autoref{fig:view-initial-web}, a view of a tool generated by \ac{spviz},
provides a very high-level view of a specific \ac{osgi} project
highlighting its architecture consisting of services, features, products, and bundle dependencies
that can be browsed to provide customizable views as shown later in the paper.

\begin{figure*}
    \centering
    \includegraphics[width=0.7\textwidth]{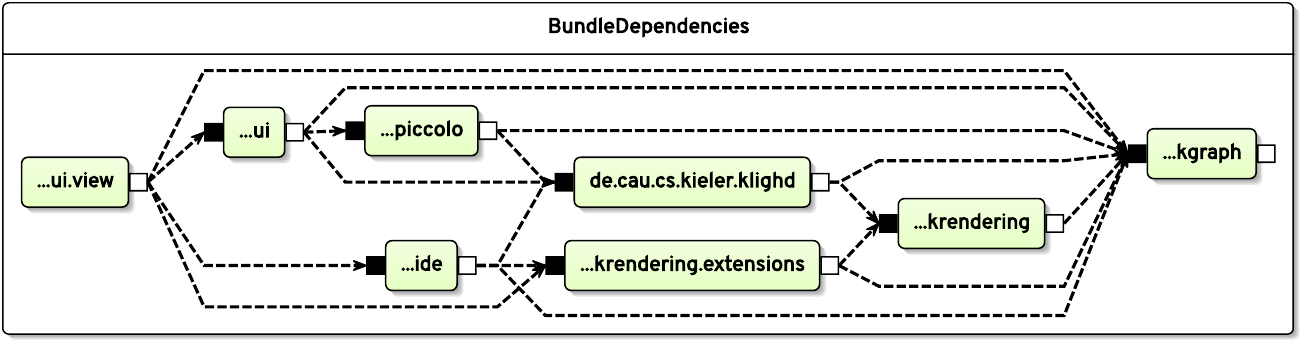}
    \caption[View of the internal bundle dependencies originating from the \emph{klighd.ui.view} bundle of the KLighD project, synthesized by the tool generated by \ac{spviz} based on structural and visual descriptions in Figures~3 and 6.]{View of the internal bundle dependencies originating from the \emph{klighd.ui.view} bundle of the KLighD\footnotemark[4]{} project, synthesized by the tool generated by \ac{spviz} based on structural and visual descriptions in \autorefs{lst:a3m-dsl-example, lst:vc3m-dsl-example}.}
    \label{fig:bundle-dep-view}
\end{figure*}
\begin{figure*}
    \centering
    \includegraphics[width=\textwidth]{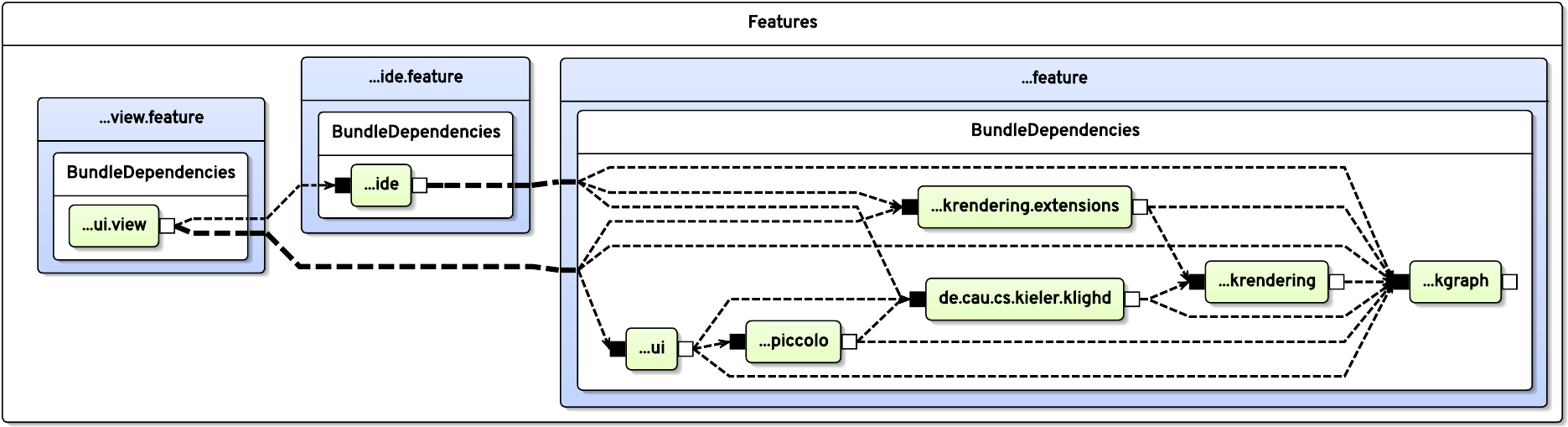}
    \caption{View on the same bundle dependency hierarchy as in \autoref{fig:bundle-dep-view}, filtered by their categorizing features.}
    \label{fig:category-connection-view}
\end{figure*}
Irrespective of the taken approach, architecture visualization tools~\cite{BoersmaL21, RentzDvH20, DuruisseauTLPG18, GeorgetTTV15} provide insights into legacy code.
However, most of them are \emph{specific} for one task or project style,
making them unusable for most other projects.
As the use of programming languages and project structures surrounding them changes over time,
new tools to work with those languages and structures must be developed.
The problem for the developers is not only to keep up with the technology,
but also with all tool support outside of rather general \acs{ide} and debugging tooling.
For each new project structure, developers will ask how project artifacts relate to each other and which hierarchies exist, to explore and explain the projects.
Alternatively to that project-specific approach,
one may use tools that support very \emph{generic} visual languages such as defined in the \acs{uml} standard.
As stated by a survey by Lange~\etal~\cite{LangeCM06},
such languages can be used and understood by many developers, architects, and other users of code.
However, such generic techniques may fail to be specific enough to describe the needs of domain experts and require too much manual effort to yield pleasing and meaningful diagrams.
Another survey by Malavolta~\etal~\cite{MalavoltaLPT+13} highlights what industry lacks in current architectural tools.
They mention \emph{support for multiple views} for architectures and adequate \emph{support for specific architectural styles and patterns}.

Users want visualizations to automatically update to changes in the actual project structure, removing the need to manually craft and update visual descriptions by hand.
Enabling a quick way to implement a new tool or tool support for software projects
to visualize their architecture offers the different use cases of visualizations
such as exploration, comprehension, and documentation to any software project.

Tools using web technologies are increasingly popular for their superior versatility and quick turnaround time with low to no installation requirements.
Interactive visualizations are no exception and should be usable on any device in any environment, for example in online documentation about systems.
Advancements in that area for example with the generic visualization framework Sprotty\footnote{
    \url{https://github.com/eclipse/sprotty/}
}, the \ac{glsp}\footnote{
    \url{https://www.eclipse.org/glsp/}
}, and a \ac{lsp} infrastructure for graphical modeling infrastructure~\cite{Rodriguez-EcheverriaIWC18}
show the feasibility of using diagrams in the web.

In the current state of the practice, for projects where no good specific tools exist and architects do not want to use the traditional way of using manually designed \ac{uml} diagrams,
they need other tool support.
Thus, the research question we address in this paper is:
\textbf{How can one create customized architecture visualization tools with minimal effort}.
In answer to that question, we here propose a \ac{dsl}-driven approach, referred to as \acf{spviz}.
\ac{spviz} uses project meta modeling with two \acp{dsl}, one to describe architectural elements and their relationships, and one to describe how these should be visualized.
Provided with such architecture and visualization descriptions, we propose to automatically synthesize a customized, project-specific visualization tool.
We have validated this approach with an open-source library, also termed \ac{spviz}\footnotemark[5]{}.
An outline of an initial view for further configuration of a concrete architecture visualization tool synthesized by \ac{spviz} is shown in \autoref{fig:view-initial-web}.
\footnotetext[4]{\url{https://github.com/kieler/klighd}} %
\footnotetext[5]{\url{https://github.com/kieler/SoftwareProjectViz/tree/spviz24}}%
\setcounter{footnote}{5} %

As proposed in our previous work~\cite{RentzDvH20} for architecture visualization specific to \ac{osgi},
the concept for that specific tool follows the \emph{modeling pragmatics} approach~\cite{FuhrmannvH10}.
Specifically, it uses the \ac{mvc} paradigm~\cite{Reenskaug79} as a guiding principle.
The approach separates the \emph{views} for exploration and navigation from the \emph{models} of the underlying projects and applies filtering and interaction possibilities as the \emph{controller}.
This previous work only works for a single architecture and therefore lacks the applicability to other architectures.
\ac{spviz} automates the process of designing such tools and adapts the concepts to arbitrary architectures.
A dependency hierarchy of an example \ac{osgi} project, where the visualization tool was generated by \ac{spviz} can be seen~in \autorefs{fig:bundle-dep-view,fig:category-connection-view}.
Here we emphasize the importance of filtering in both the architectural description and visualization tooling to not show everything at once, but focus on individual components of a system and their context.
The figures are interesting and readable, because they do \emph{not} show many details, but filtered views,
here for the investigation of the project-internal dependency hierarchy of a specific bundle~artifact.

\emph{Outline:} In \autoref{sec:visuals} we recapitulate how the visualization and interaction style proposed in our previous work~\cite{RentzDvH20} works for \ac{osgi} projects.
Next, we cover the \textbf{main contributions}:
\begin{itemize}
    \item We present the \ac{spviz} approach via two \acp{dsl} describing arbitrary architectures and their visualization as a generalization of the \ac{osgi} visualization tool in \autoref{sec:dsls}.
    \item We propose how to automatically generate a complete visualization tool, as well as template code for a model generator akin to the \ac{osgi} visualization using the \acp{dsl} based on example project architectures in \autoref{sec:developing}.
    \item We illustrate the \acp{dsl} using example projects, both open source and ones from industrial partners, and present the generated tools with their feedback in \autoref{sec:evaluation}.
\end{itemize}
Finally, we compare related work in \autoref{sec:related}, discuss further directions in \autoref{sec:discussion} and conclude in \autoref{sec:conclusion}.

%%%%%%%%%%%%%%%%%%%%%%%%%%%%%%%%%%%%%%%%%%%%%%%%%%%%%%%%%%%%%%%%%%%%%%%%%%%%%%%%%%%%%%%%%%%%%%%%%%
% -------------------------------------- The Main Sections ------------------------------------- %
%%%%%%%%%%%%%%%%%%%%%%%%%%%%%%%%%%%%%%%%%%%%%%%%%%%%%%%%%%%%%%%%%%%%%%%%%%%%%%%%%%%%%%%%%%%%%%%%%%

\section{Visualizing Specific Project Architectures}
\label{sec:visuals}
\begin{figure}
    \centering
    \includegraphics[width=\columnwidth]{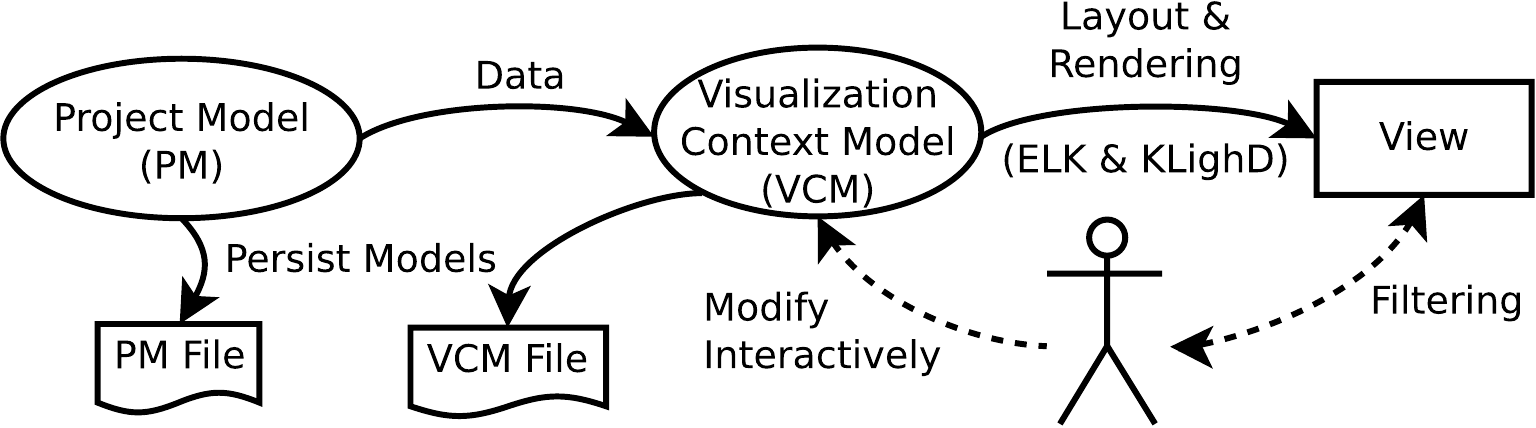}
    \caption{The usage process of the view tools with its core, the \acl{vcm}. It is used to configure and filter views for later reuse. Solid arrows depict data flow, dashed ones the interaction paths to control the \acs{vcm}. Adapted from Rentz~\etal~\cite{RentzDvH20}.}
    \label{fig:visualization-context-interaction}
\end{figure}
\begin{figure}
    \centering
    \includegraphics[width=0.4\columnwidth]{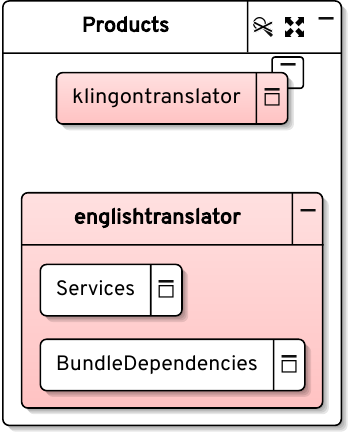}
    \caption{Collapsed (klingontranslator) and expanded (englishtranslator) products. Expanded product shows filtered views (services, bundle dependencies), here in their current collapsed form.}
    \label{fig:containment-artifact-views}
\end{figure}
\begin{figure*}
    \centering
    \includegraphics[width=0.85\textwidth]{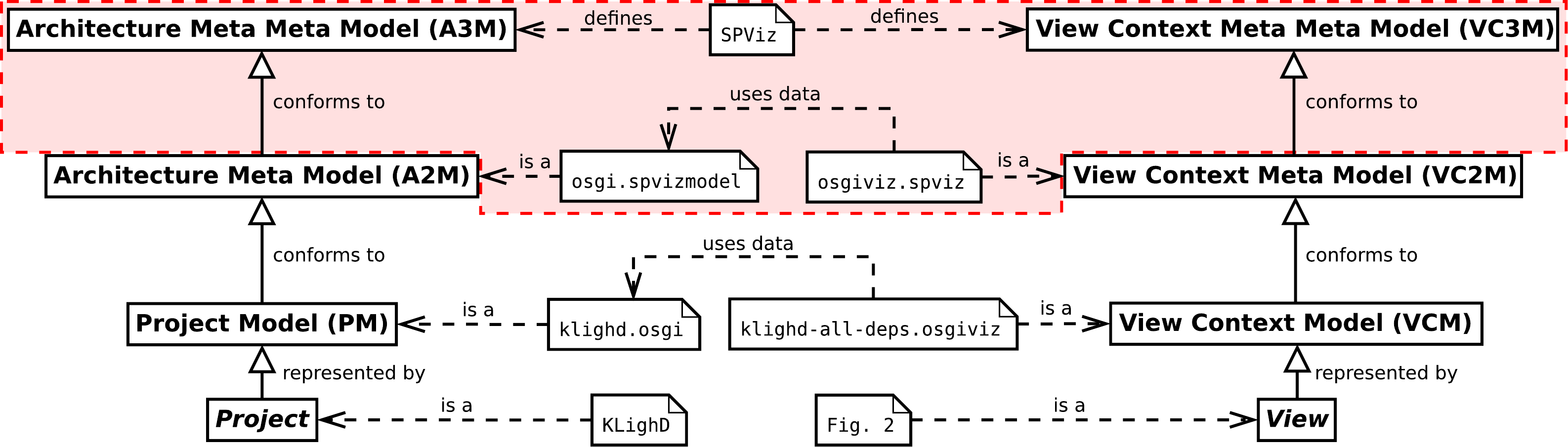}
    \caption{The meta modeling hierarchy of \acs{spviz}. The shaded top is our proposed abstraction, the center contains examples for the different models.}
    \label{fig:metamodeling-hierarchy}
\end{figure*}
In our previous work~\cite{RentzDvH20}, we presented a visualization tool specific to the \ac{osgi} architecture to aid developers understand legacy \ac{osgi} projects and to document actively developed systems.
The use of the diagrams, according to the previous paper,
allows users to move from manually drawn diagrams such as \ac{uml} to automatically created diagrams
that are useful for system documentation purposes.
The visualization tool utilizes the tooling of the \ac{klighd} framework~\cite{SchneiderSvH13} with automatic layout by the \ac{elk}\footnote{\url{https://www.eclipse.org/elk/}}.
\ac{klighd} provides a visualization of \ac{osgi} projects given a \emph{model synthesis},
which is implemented in that tool.
This visualization uses node-link diagrams to represent structural relationships between architectural artifacts.
This follows the \emph{graph-based} visualization technique, a term coined by Shahin~\etal in a systematic literature review of software visualization~\cite{ShahinLAB14}.
This section summarizes the concepts and visualization techniques,
the later sections generalize these concepts for arbitrary architectures.

The \emph{\ac{vcm}}, depicted in \autoref{fig:visualization-context-interaction}, is the central model for interaction with the tool by modifying and filtering views to be reusable for documentation purposes in evolving (software) projects.
This model is entirely hidden from the user and only modified by interaction with the UI.
The \emph{\ac{pm}} contains the extracted data of a concrete project, here for an \ac{osgi} project.
It is the data source of the \ac{vcm} and describes the project at its state in time when the \ac{pm} was generated.
The \ac{pm} conforms to the meta model of the OSGi architecture,
therefore we also call this the \emph{\ac{a2m}}.
It models all possible \acp{pm} for \ac{osgi} projects.
When initially visualizing a \ac{pm}, a \ac{vcm} is created. Together with the model synthesis and \ac{klighd},
this allows visually browsing different views sensible in the \ac{osgi} environment.

\autoref{fig:view-initial-web} shows the view of such an initial \ac{vcm} as it is initially created for any \ac{osgi} project.
Interacting with the view via the options, filters, or the \ac{ui} will change the \ac{vcm} directly to reflect the currently shown and connected elements.
\autoref{fig:bundle-dep-view} is such a configured view investigating the \ac{klighd} framework,
which also uses \ac{osgi} as its project architecture. The view is configured to focus on the bundle dependencies view to show all bundles
that are directly or indirectly required by its \emph{klighd.ui.view} bundle.
All connections of the project
as defined in the \ac{a2m} and configured to be shown in the meta model for the \ac{vcm}, the \ac{vc2m}
can be interactively added or removed to show any hierarchy.

Filtered views based on \eg \ac{osgi} products can be browsed as shown in \autorefs{fig:category-connection-view, fig:containment-artifact-views}.
In \autoref{fig:containment-artifact-views} the products show child views for their services and for their bundle dependencies, which can be expanded to browse only the service artifacts/bundles used in that product.

By interacting with the views with the mouse, the \ac{vcm} and with that the view can be modified.
Overall, the user can
\begin{itemize}
    \item (un-)focus and expand/collapse views,
    \item show/hide all collapsed artifacts,
    \item connect the connected artifacts of an artifact following a connection once, or for all artifacts at once,
    \item remove all connections of an artifact,
    \item undo/redo the last action or reset the view to its default,
    \item filter to show or hide the individual artifact views,
    \item filter artifacts based on their IDs,
    \item export a \ac{vcm} to document specific parts,
    \item and more standard browsing interactions.
\end{itemize}
Of these, showing/hiding the collapsed artifacts, connecting artifact connections recursively, and removing all connections from an artifact are not in the previous work and newly implemented in \ac{spviz} based on feedback by our industrial partner.

The collapsed and expanded artifacts are always shown beneath each other in two separate areas, as shown in \autoref{fig:containment-artifact-views}.
On top the collapsed artifacts are packed together, where the tool features a search and filter of the artifacts to be inspected further.
On the bottom the expanded artifacts are shown with their filtered artifact views or their connections to other artifacts in the view as in \autoref{fig:bundle-dep-view}.
Furthermore, in \autoref{fig:bundle-dep-view} the collapsed artifacts are also hidden and the interactive buttons are deactivated, to allow the export of such a \ac{vcm} as part of the documentation for that project without cluttering the view.

In \ac{spviz}, the colors of all artifacts are picked in a way to have a low saturation to have a good contrast to black text, while trying to not collide with the colors of other artifacts.
This is done by choosing the hue of the color based the low-discrepancy sequence on remainders mod 1 of integer multiples of the golden ratio~\cite{SchretterKD12}, to have consistent colors with good contrast between them, no matter how many different artifacts are used in the \ac{a2m}.
The colors could also be manually picked by the visualzation tool designer to match known styles for the specific architecture.

Overall, this previous work~\cite{RentzDvH20} can be used for \ac{osgi} projects,
but lacks usage for any other architecture.
We now further generalize this visualization tool and make it applicable to arbitrary architectures.

\section{The \acs{spviz} \acsp{dsl} as Architecture \\and View Meta Meta Models}
\label{sec:dsls}
To allow domain experts to conceptualize a visualization for software projects following arbitrary \acfp{a2m},
we define meta meta models to describe the general structure of a software architecture and an abstract way of visualizing that architecture.
We illustrate the meta modeling hierarchy of visualization tools relative to our previous work~\cite{RentzDvH20} with our proposed abstraction in \autoref{fig:metamodeling-hierarchy}.
\autoref{sec:visuals} explains the project, view, and their respective \acf{pm} and \acf{vcm}.
We refer to the meta model for describing the architecture the \emph{\acf{a2m}},
and the meta model for the possible types of shown connections and views the \emph{\acf{vc2m}}.
This section introduces two \acp{dsl} for defining such \acp{a2m} and \acp{vc2m},
thus making the \acp{dsl} themselves an \emph{\ac{a3m}} and a \emph{\ac{vc3m}}.

\subsection{The Architecture Meta Meta Model}
\label{sec:project-model}
Domain experts can describe a project architecture using our definition of the \emph{\acf{a3m}} \ac{dsl}.
The grammar is defined using the Xtext~\cite{EfftingeV06} framework
and shown in \ac{ebnf} in \autoref{lst:a3m-dsl-grammar}.

All project structures are different in their concrete realization in the sense of which files and which configurations define the underlying project.
However, in an abstract sense projects always contain different \emph{artifacts} and \emph{references} between these artifacts.
\emph{Artifacts} can be coarse- or fine-grained parts of a software system such as entire products, features, classes, or even statements,
which may refer to other artifacts.
\emph{References} can be further specialized into \emph{connections},
\eg dependencies connecting different artifacts,
and \emph{containments},
\eg some product artifact containing a set of packages.
We define these components as the \ac{a3m}.
The \ac{a3m} can be applied to most architectures to show how all their different artifacts relate to each other.
This concept is comparable to other meta models used for \ac{mde} such as the \ac{mof}~\cite{OMG-MOF19},
in a simplified version.
We do not need its relations special to object orientation,
such as class inheritance to describe arbitrary software architectures.
We have found
that the \ac{a3m} suffices to describe interrelations.
In fact, class inheritance can be modeled with our model through the use of references from a class artifact to itself.
This is due to the \ac{a3m} being self-describing, just as the \ac{mof}.
Furthermore, the \ac{a3m} restricts the designer to design their architecture in a minimal way and to restrict it to a unique way for later code generation.

\newsavebox{\projectModelGrammar}
\begin{lrbox}{\projectModelGrammar}%
    \begin{lstlisting}[style=EBNF] 
SPVizModel = "package" Name
    "SPVizModel" Name "{"{Artifact}"}" 
Artifact = Name ["{"{Reference}"}"]
Reference = Containment|Connection
Containment = "contains" Artifact
Connection = Name "connects" Artifact
Name = ? Letters, numbers, full stop ?
             
\end{lstlisting}
\end{lrbox}

\begin{figure}[t]
    \centering
    \subfloat{\usebox{\projectModelGrammar}}
    \caption{\Ac{a3m} \acs{dsl} grammar.}
    \label{lst:a3m-dsl-grammar}
\end{figure}

\newsavebox{\simpleOsgiModel}
\begin{lrbox}{\simpleOsgiModel}%
    \begin{lstlisting}[style=spvizmodel]
// generate code into this package
package de.cau.cs.kieler.spviz.osgi

// name of the project structure is OSGi
SPVizModel OSGi {
  // the artifacts the project contains
  Feature {
    // features give structure to their bundles
    contains Bundle
  }
  Bundle {
    // bundles may connect to other bundles
    // as a connection called "Dependency"
    Dependency connects Bundle
    // services are defined within bundles
    contains ServiceInterface
    contains ServiceComponent
  }
  // service components can require and provide
  // other service interfaces. Inverting the
  // provision for a consistent "requires" direction
  ServiceInterface {
    ProvidedBy connects ServiceComponent
  }
  ServiceComponent {
    Required connects ServiceInterface
  }
}
\end{lstlisting}
\end{lrbox}

\begin{figure}[t]
    \centering
    \subfloat{\usebox{\simpleOsgiModel}}
    \caption{Example \acs{a3m} \acs{dsl} usage.}
    \label{lst:a3m-dsl-example}
\end{figure}
\begin{figure}[t]
    \centering
    \includegraphics[width=0.9\columnwidth]{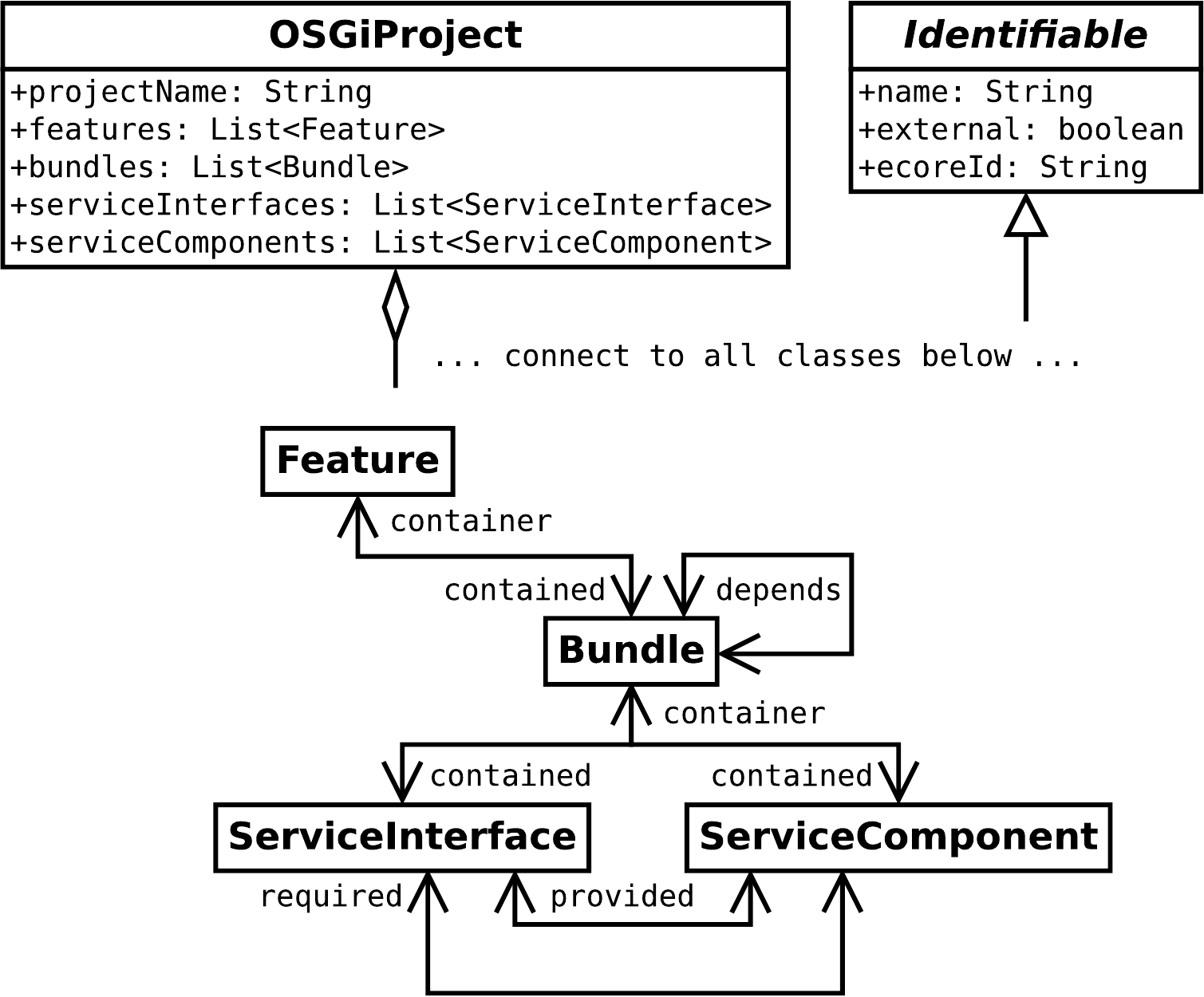}
    \caption{Class structure as generated for the example \ac{osgi} \ac{a2m}.}
    \label{fig:osgi-class-diagram}
\end{figure}
\begin{figure}[t]
    \centering
    \includegraphics[width=0.97\columnwidth]{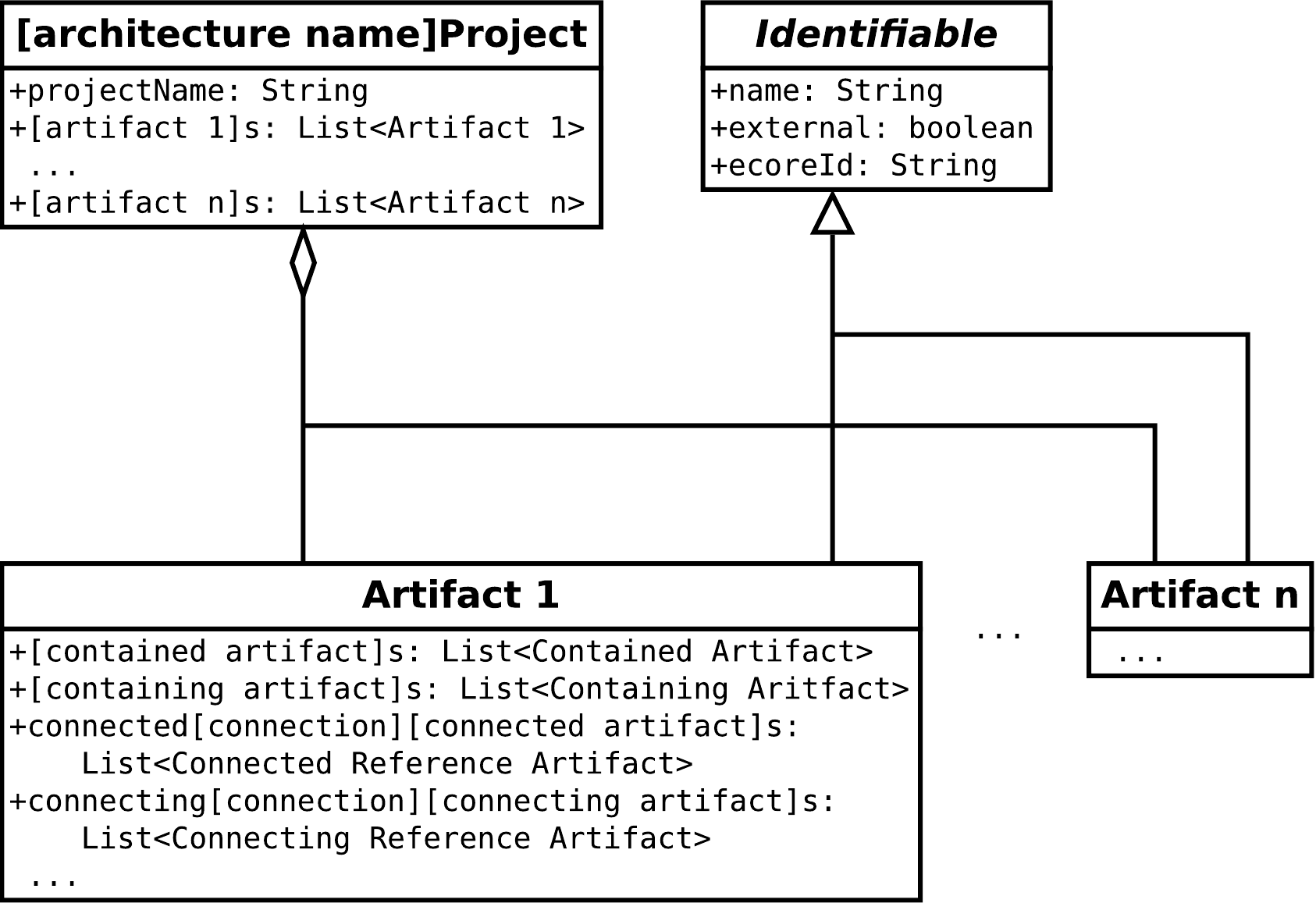}
    \caption{Abstract class structure of the generated code for the \acs{a2m}.}
    \label{fig:project-model-class-diagram}
\end{figure}
\autoref{lst:a3m-dsl-example} shows an example use of the \ac{a3m} \ac{dsl} to describe the \ac{osgi} architecture \ac{a2m}.
This example intentionally re-visits the \ac{osgi} architecture to show that the previous tool~\cite{RentzDvH20} can entirely be generalized.
The information in the example consists of the name, artifacts, their hierarchy, and connections.
The resulting meta model describes coarse- and fine-granular artifacts of the \ac{osgi} architecture modeling bundle dependencies and service hierarchies.
Blocks within the \emph{SPVizModel} block define the artifacts that the architecture contains, here \emph{features}, \emph{bundles}, \emph{service interfaces}, and \emph{service components}.
These artifacts are structured according to the artifacts that they contain.
In this example, the features are structured by the bundles that they contain.
This artifact structure can be used to automatically generate a class hierarchy,
which is~part of the subsequent code synthesis.
A corresponding class diagram with containments modeled as references is shown in \autoref{fig:osgi-class-diagram}.
\autoref{fig:project-model-class-diagram} shows the pattern for the class structure for any \ac{a2m} in general.
Every artifact type gets its own class and the connections and containments are handled via reference lists.

\ac{pm} instances of this OSGi \ac{a2m} describe information of the structure of concrete projects to model dependencies between bundles from the project itself and external ones.
One can also get an insight into otherwise more obscure connections of different service components, as defined in the service layer of the OSGi specification~\cite{Osgi20}, in the specification with the name \emph{service objects}.

This example \ac{osgi} \ac{a2m} taken from the \ac{a3m} \ac{dsl} is now comparable to the manually described one from \autoref{sec:visuals}.
However, only using the \ac{a3m} is insufficient to define how different views should be configured and filtered.
For that, we define the \emph{\ac{vc3m}} as another \ac{dsl}.

\subsection{The View Context Meta Meta Model}
\label{sec:view-model}
The \emph{\acf{vc3m}} makes it possible to describe which of the artifacts and their connections from the \ac{a3m} should be visualized in different views.
Typically, not all possible connections should be shown in any view, and not all artifacts of the same type should be in the same view part.
Just showing everything at once is typically not the best visualization for project structures,
but filtered subsets are.

\newsavebox{\viewModelGrammar}
\begin{lrbox}{\viewModelGrammar}%
    \begin{lstlisting}[style=EBNF] 
SPViz = "package" Name
    "import" URI
    "SPViz" Name "{"
        {View}
        {ArtifactShows}
    "}" 
View = Name "{"
    {ShownElement}
    {ShownConnection}
    {ShownCategoryConnection}
"}"
ShownElement = "show" Artifact
ShownConnection = "connect" Connection
ShownCategoryConnection = "connect" Connection
    "via" Artifact {">"Artifact} "in" View
ArtifactShows = Artifact "shows" "{"
    {ArtifactView}
"}"
ArtifactView = View "with" "{"
    {Artifact "from" Artifact {">"Artifact}}
"}"          
\end{lstlisting}
\end{lrbox}

\begin{figure}[t]
    \centering
    \subfloat{\usebox{\viewModelGrammar}}
    \caption{\acs{vc3m} \acs{dsl} grammar, referring to the \ac{a3m} \ac{dsl} grammar in~\autoref{lst:a3m-dsl-grammar}.}
    \label{lst:vc3m-dsl-grammar}
\end{figure}

\newsavebox{\simpleOsgiViewModel}
\begin{lrbox}{\simpleOsgiViewModel}%
    \begin{lstlisting}[style=spviz, escapechar=!]
// generate code into this package
package de.cau.cs.kieler.spviz.osgiviz
// refer to the "OSGi" model above
import "osgi.spvizmodel"

// this visualization is called "OSGiViz"
SPViz OSGiViz {
  // the available views for OSGiViz
  // this is the view for services
  Services {
    // show the service hierarchy with all
    // its artifacts and connections
    show    OSGi.ServiceInterface
    show    OSGi.ServiceComponent
    connect OSGi.ServiceInterface.ProvidedBy
    connect OSGi.ServiceComponent.Required
  }
  // view for bundles and their dependencies
  BundleDependencies {
    show    OSGi.Bundle
    connect OSGi.Bundle.Dependency
  }
  // view of features, for filtering
  Features {
    show OSGi.Feature
    //connect features via their bundle dependencies
    connect OSGi.Bundle.Dependency via OSGi.Feature
        in BundleDependencies
  }

  // features can show filtered artifact views,
  OSGi.Feature shows {
    // inner views connected as defined above
    BundleDependencies with {
      // only bundles contained in the feature
      OSGi.Bundle from OSGi.Feature!\textbf{>}!OSGi.Bundle
    }
  }
}
\end{lstlisting}
\end{lrbox}

\begin{figure}[t]
    \centering
    \subfloat{\usebox{\simpleOsgiViewModel}}
    \caption{Example \ac{vc3m} \acs{dsl} usage, referring to the example in \autoref{lst:a3m-dsl-example}.}
    \label{lst:vc3m-dsl-example}
\end{figure}

Therefore the \ac{vc3m} allows the architects to further structure their  visualization.
Its grammar defined by an Xtext \ac{dsl} is shown in \autoref{lst:vc3m-dsl-grammar}.
The \ac{vc3m}, connecting to an existing \ac{a3m}, describes \emph{views}.
These views refer to the \ac{a3m} and define which artifacts and which of their connections can be visualized within them.
This allows for configurability, to have views specifically intended to show one kind of connection between artifacts, or multiple connection types for broader overviews.
Furthermore, the \ac{vc3m} makes a configuration of how the artifacts themselves should be visualized possible.
As artifacts from the \ac{a3m} may contain other artifacts,
we allow the definition of \emph{artifact views} as views filtered to the context of this parent artifact.
Artifact views can be used especially for artifacts that have the purpose of organizing other artifacts,
such as bundle categories as subsets for bundles in the \ac{osgi} specification~\cite{Osgi20}.
The artifact views refer to the views defined above with their artifact and connection types
and filter the artifacts to the context of this artifact view's parent artifact.
Finally, the \emph{category connections} are another way of organizing and filtering views.
They are part of views and show the implicit connections between categorizing artifacts
that are in a relation to each other via some contained artifact type.
Category connections are defined by the connection they represent, the artifact hierarchy following the parent-child hierarchy to get from the connected artifacts to the defined connection, as well as the inner view to connect to.
The following example will further illustrate this concept.

\begin{figure*}
    \centering
    \includegraphics[width=0.8\textwidth]{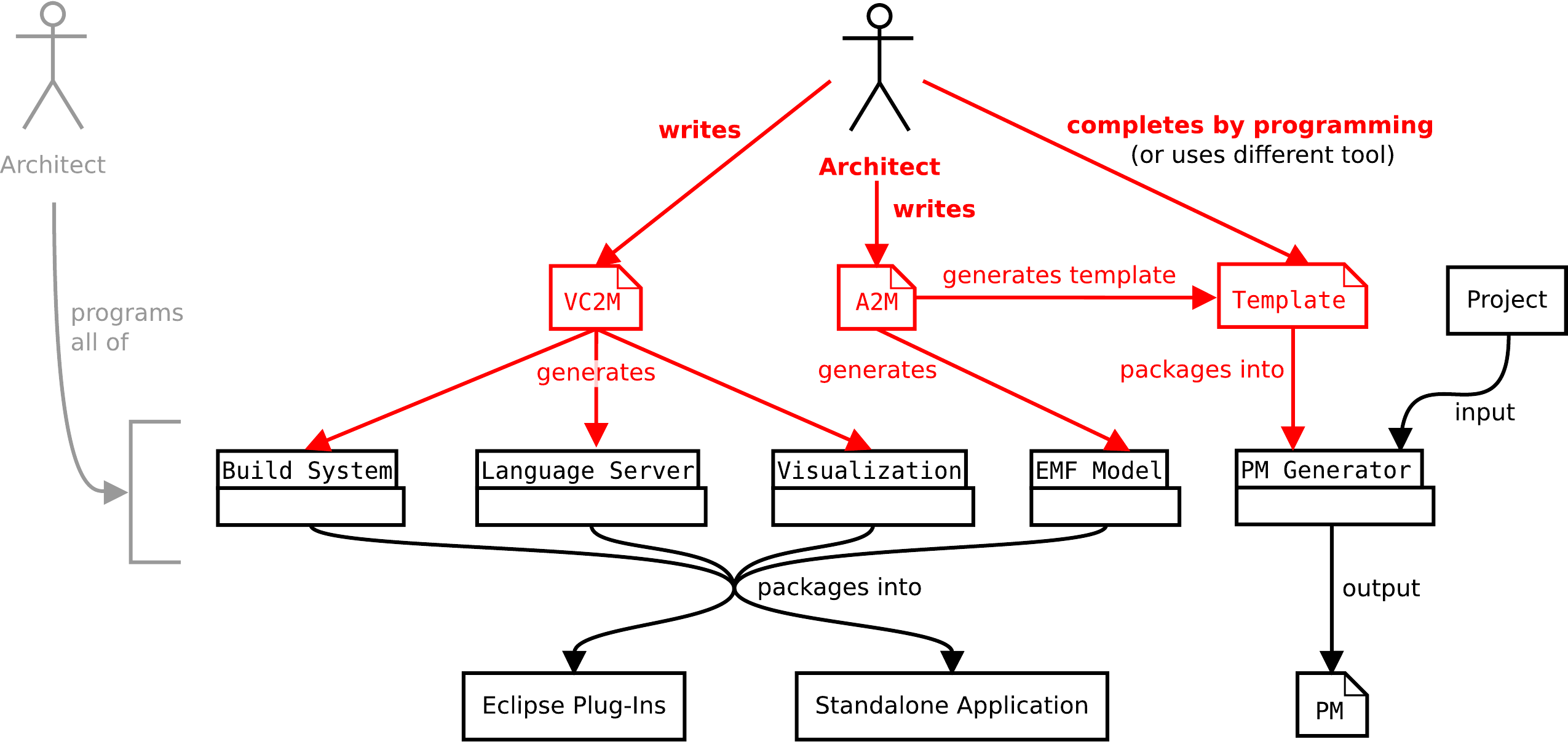}
    \caption{Traditional (left, gray) and proposed (red) process for developing a new software project visualization tool. \acs{a2m} and \acs{vc2m} are written in the \acp{dsl} proposed here. Boxes represent files, software packages, and applications in \ac{uml} style.}
    \label{fig:developing-artifacts}
\end{figure*}
Continuing the \ac{osgi} example, \autoref{lst:vc3m-dsl-example} shows a possible use of the \ac{vc3m} \ac{dsl} to describe a \ac{vc2m} for the \ac{osgi} \ac{a2m}.
The configuration matches and extends the \ac{osgi} visualization described in \autoref{sec:visuals}, thus showing that the previously manually designed \ac{osgi} visualization can be generalized using our \acp{dsl}.
The example defines a new visualization for the \ac{osgi} architecture called \emph{OSGiViz} and defines what views can be shown in general, as well as how artifacts can reuse these views to filter views down into a sensible context.
The views called \emph{services} and \emph{bundle dependencies} clarify that the artifacts and the connections related to them from the underlying \ac{osgi} model should be shown.
An example view of the bundle dependencies can be seen in \autoref{fig:bundle-dep-view}.
The view named \emph{features} shows an overview of all possible features and a category connection.

Through configuration of an \emph{artifact view}, the features displayed in their overview show filtered views specific to the individual features defined within an \ac{osgi} project.
This artifact view defines the filters for a bundle dependencies view.
It defines how the bundles that should be shown in a bundle dependencies view as defined above are filtered.
The \ac{dsl} allows referring to the requested artifacts by chaining the artifacts together via their \emph{contains} relation of the \ac{a3m}.
In this example it means that all bundles are shown that are listed in the feature's child bundles.
This way, whereas a general view for bundle dependencies would show all bundles as they are used and defined in the whole project,
the feature-specific artifact view for bundle dependencies shows the filtered view, only with bundles relevant for the feature.
\autoref{fig:category-connection-view} illustrates this, as each feature only shows the bundles in their bundle dependencies view that the feature contains.

The \emph{category connection} defined in the features view makes it possible to show a relation between features,
although the \ac{osgi} \ac{a2m} does not define any direct connections for features.
Because features contain bundles, which themselves define the bundle dependencies connection, this category connection enables showing the relation between the features in terms of their bundle dependencies.
That is, a connection between features will be shown, if a bundle contained in one feature has a connection to a bundle contained in another feature.
\autoref{fig:category-connection-view} illustrates this.
Here the \emph{view} and \emph{ide} features have a shown dependency,
because the \emph{ui.view} bundle within the \emph{view} feature has a dependency to the \emph{ide} bundle within the \emph{ide} feature,
as also seen in \autoref{fig:bundle-dep-view}.
Edges are drawn thicker if they bundle multiple dependencies into a single edge.
This is an important filtering view that does not require more effort in the later \ac{pm} extraction, but just a single line in the \ac{vc2m}.

\section{Project Visualization Tool Synthesis with \ac{spviz}}
\label{sec:developing}
\autoref{fig:developing-artifacts} shows the proposed development process for software architects who want to apply the visualization technique from our previous work~\cite{RentzDvH20} to their project.
They need to describe the architecture and its visualization as described in \autoref{sec:dsls}
and extract information about the real artifacts of the project on their file system into a \ac{pm}.
To design the models, questions such as \enquote*{What connections and hierarchies in the code should be made visible?} and \enquote*{What kinds of artifacts define and order these?} need to be answered.
This first layer of modeling a visualization and the architecture is the main task of an architect when using \ac{spviz}.
Traditionally, the architect does not have access to this first layer and needs to develop all code in the second layer manually or use an entirely different approach.
This traditional way was done in our previous work~\cite{RentzDvH20}, showing the simplification the \acp{dsl} provide compared to that process.

\subsection{Visualization Framework Generated from the \acsp{dsl}}
\label{sec:developing-generated}
Our implementation of the \ac{spviz} approach separates the generated visualization into multiple parts and uses \ac{ide} support
and the code generation features of Xtext for \acp{dsl} for that purpose.
This code generation is the second layer of steps in \autoref{fig:developing-artifacts} and will be automatically executed when developing the \acp{dsl} in Eclipse.
Once the user is finished designing their \ac{a2m}, \ac{spviz} will create an \ac{emf} model of that \ac{a2m} following the class structure in \autoref{fig:project-model-class-diagram}.
Such a model contains a single project class containing lists of all artifacts defined for that project.
For the \ac{osgi} example from above, this results in the project class \emph{OSGiProject} as in \autoref{fig:osgi-class-diagram} containing lists of its features, bundles, service interfaces, and service components.
Every artifact then has lists that refer to all artifacts they are in a containment or connection relation to.
\Ie each artifact refers to the source or target artifact for each connection it is used in,
as well as the parent or child artifact for each containment it is used in.
More specifically, bundles refer to their contained service interfaces and components,
to the features they are contained by,
as well as the bundles they refer to and are referred by the dependency connection.
That means that every relation can be followed both ways in this model and any change to any of the lists will automatically update its opposing list by \ac{emf}.

The second part created by \ac{spviz} is a template for a \ac{pm} generator.
The template is a complete program with a dependency on the generated \ac{emf} model of the \ac{a2m}.
It comes with a Maven build, which can bundle it into an executable.
The template contains a file \emph{ReadProjectFiles.java} which is missing the architecture-specific extraction of data from the project's sources.
It provides methods to create and connect all artifacts as defined in the \ac{a2m}
as well as a checklist of all artifacts, connections, and containments that need to be extracted in the generator.
Examples described in \autoref{sec:evaluation} implement this template to show its feasibility.
Alternatively, an extractor can be a separate program or tool.
Our \ac{api} requires an \ac{emf} model instance following the class structure of \autoref{fig:project-model-class-diagram} in the open \ac{xmi} format proposed by OMG~\cite{OMG-XMI15},
which can be implemented by any tool.
For more information about the \ac{xmi} model \ac{api}, see the \ac{emf} book~\cite{SteinbergBP+09}.
This open format allows other code mining and reverse engineering tools to work together with our \ac{a2m} and therefore with the visualization tool.

Because the \ac{a2m} and its generator can be used isolated from the \ac{vc2m}, this allows users to write different \acp{vc2m} for the same \ac{a2m}.
This is useful so that a single model generator can create a model of the current state of a project, which can be visualized using different \acp{vc2m}.

Once the user is finished designing their possible views in the \ac{vc2m}, \ac{spviz} will create four more modules:
the \ac{emf} model and code to support that \ac{vc2m}
that yields a visualization using the KLighD framework~\cite{SchneiderSvH13},
a language server to enable viewing KLighD diagrams in web environments~\cite{Rentz18},
as well as a Maven build system.
This can be packaged together either as Eclipse plug-ins or as a standalone application to be used together with the KLighD \acs{cli}~\cite{Fricke21}\footnote{
    Available on GitHub: \url{https://github.com/kieler/klighd-vscode}
}.

As a different concept, one could think of visualizing systems just using the \acp{dsl} without code generation,
thus having a single generic \ac{spviz} tool that could handle all visualizations, given the \acp{dsl}.
However, that would complicate writing a model generator, as it could not programmatically depend on the \ac{a2m} model code, but would require more complex meta-parsing of models and it would ignore the benefits of the \ac{mde} workflow with the \ac{emf} models.
Furthermore, the tool itself would need the \acp{dsl} to run and parse the configurations, making the tool less concise for its use case.
However, such a generic tool would remove the need to generate, build, and execute the built tool first from the \acp{dsl} and allow the direct execution of such a tool, thus creating a tradeoff.
We opted for the generation approach, mainly to make the subsequent implementation/adaption of the model generator simpler.

\subsection{View Your Visualization Anywhere}
\label{sec:integration}
\ac{spviz} is designed to work with any project architecture, causing the range of such an architecture's main development \ac{ide} to be pretty much any \ac{ide}.
The tool support for development in any language should try to follow the \ac{ide}, as for example requested by Charters \etal~\cite{ChartersTM03} for visualization tools or presented in tools such as VisUML~\cite{DuruisseauTLPG18}.
As there is no widely-adopted framework or \ac{api} to add diagrams in many \acp{ide} yet, such an integrated support is not feasible.
However, the use of web tools and tools developed with web technology in mind is growing and allows for an easy access.
Therefore, we also make use of web technologies for the configuration and deployment of architectural views.

While individual deployment of tools such as the individual tools generated by \ac{spviz} can vary between projects, they can be integrated and configured into of a largely automated process after their initial setup to view the models in the web or some \acp{ide} directly.

The initial setup of the individual visualization tool is supported mainly in the Eclipse \ac{ide}.
This includes the development of both \acp{dsl} and the implementation/adaption of a model generator for the \ac{pm}.
From there, most other use cases can be automated and integrated into existing web documentation.
We think that after the initial setup, the tool is best used by first integrating the model generator into the \ac{ci}/\ac{cd} chain of the project.
The model generator template has a Maven build, which will create an executable to call in \ac{ci}/\ac{cd} for every iteration where it should update the documentation.
Architects can use the generated \ac{pm} as well as the visualization tool to configure some specific views and to put their corresponding \acp{vcm} into the documentation.
Here the architects are free to configure the views anywhere the final visualizations can be displayed as well.
That is, the diagrams can currently be shown in the Eclipse \ac{ide}, in \ac{vscode}, or embedded in web sites (see \autoref{fig:view-initial-web}) by using KLighD or the KLighD \ac{cli}.
The view configuration of that \ac{pm} can be stored as its \ac{vcm} and deployed to the documentation, making it possible to update the \ac{pm} generated by the \ac{ci}/\ac{cd} to always have most up-to-date documentation without the need to adapt the \ac{vcm}.

As another use case, the visualization applications and \ac{pm} generator designed for the architecture can also be distributed to other developers of the project for exploration outside of the pre-configured documentary views.

\section{Preliminary Evaluation and Validation}
\label{sec:evaluation}
We evaluate our proposed concept in two ways.
First, we show its flexibility and usability for diverse project architectures by realizing four different \acp{a2m} and \acp{vc2m} via the \acp{dsl}, motivated by open source projects and projects developed by our industrial partner.
For each resulting tool, we evaluate the tool usability with these open source and industrial projects
and do a quantitative analysis on the effort reduction using \ac{spviz} for the \ac{osgi} example.
Second, we evaluated different user stories for different user groups of \ac{spviz} and asked two users of different projects of our industrial partner for feedback on their goals with the generated project visualizations and their successes and criticisms.

\subsection{Testing with Real-World Examples}
\label{sec:evaluation-testing}
We answered the design questions as mentioned in \autoref{sec:developing} for four different project architectures and modeled the \acp{a2m} and \acp{vc2m} accordingly.
As some examples are rather specific on the project configurations, \eg being for a specific build and dependency system with a specific \ac{di} framework, they do not directly apply to most other projects.
However, they are easily configurable and combinable, so that tools working for other architectures with their specific use cases are built quickly.
The models and generators for the examples can be found in the \ac{spviz} examples repository\footnote{
    \url{https://github.com/kieler/SoftwareProjectViz-examples/tree/spviz24}
}.

\subsubsection{\acs{osgi}}
\label{sec:examples-osgi}
For the OSGi visualization, the created models aim to visualize dependencies within the \emph{module layer} and service relations within the \emph{service layer} of the \ac{osgi} specification~\cite{Osgi20}.
The example, which is slightly extended compared to the \ac{osgi} example from \autorefs{lst:a3m-dsl-example, lst:vc3m-dsl-example}, also uses \emph{products} to further organize the individual components
and allows the view of service connections in relation to their parent bundle artifacts as category connections.
This example architecture visualization was designed as a first proof of concept.
It indicates that the interactive visualizations can in fact be generalized from the implementation of our previous work~\cite{RentzDvH20} and result in equivalent views conveying the information that the architecture-specific visualization does.
The \ac{pm} generator for \ac{osgi} projects scans through a repository and parses the data of \emph{META-INF/MANIFEST.MF} files for bundles, \emph{feature.xml} files for features, and \emph{*.product} files for products. Furthermore, it parses \emph{OSGI-INF/} folders and the Java files for definitions and connections of service components and interfaces and creates an \ac{osgi} \ac{pm} based on that data for any \ac{osgi} project.

\begin{figure*}
    \centering
    \includegraphics[width=\textwidth]{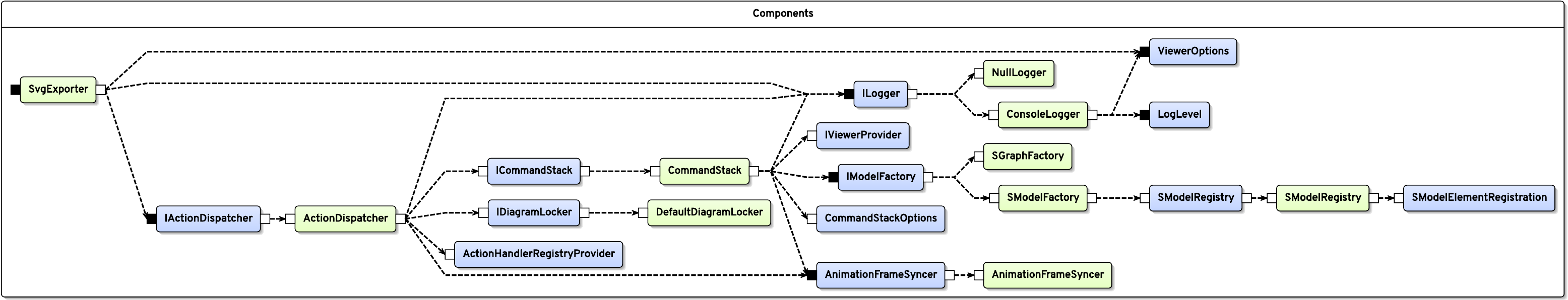}
    \caption{The interfaces and resolved classes providing those interfaces for the \emph{SvgExporter} class as they are pre-configured by Sprotty's main repository.}
    \label{fig:view-sprotty-exporter}
\end{figure*}
This example was verified with a project from our industrial partner, as well as the \ac{klighd} and Semantics frameworks of the KIELER\footnote{\url{https://github.com/kieler}} project. The partner project consists of 144 bundles plus 109 additional dependent bundles, as well as 285 service artifacts. The \ac{klighd} and Semantics frameworks consist of 25 plus 196 bundles and 166 plus 144 bundles, respectively.
% Industrial partner project 1: 144/253 (internal/overall) bundles, 285 service artifacts
% Industrial partner project 2: 109/228 (internal/overall) bundles, 196 service artifacts
% KlighD: 25/221, no services
% semantics: 166/310, no services
Example views of this are shown in \autorefs{fig:bundle-dep-view,fig:category-connection-view}.

To compare the effort of a manual architecture-specific implementation and the \ac{spviz} approach,
we compare the manually written \ac{loc} using the \ac{spviz} approach for the \ac{osgi} example and the code generated from our tool.
We compare to the generated code instead of the manual implementation because of slight differences and improvements between the tools and similar \ac{loc} in the generated and manual tools.
The \ac{a2m} and \ac{vc2m} descriptions using the \acp{dsl} for this \ac{osgi} example have a combined 76 \ac{loc}
and the \ac{pm} generator template was extended by 466 \ac{loc} in Java to extract all \ac{osgi}-specific artifacts, their connections, and containments of any such \ac{osgi} project.
The generated project consists of 318 \ac{loc} of \ac{emf} model code and 8654 \ac{loc} of Xtend and Java code that would have to be written manually without \ac{spviz}, plus the 466 \ac{loc} of above to make the generator template functional.
These numbers do not include the 25\,000 lines of Java code that are further generated from the generated \ac{emf} models, as they do not add to manual work in either setup.
Comparing the effort on a \ac{loc} basis, this yields a reduction in the manually written \ac{loc} of $ 1 - \frac{76 + 466}{318 + 8654 + 466} = 94.3 \% $.
The effort is further reduced because the generated tool comes with a release engineering configuration
that builds and packages the generated tool as an Eclipse plugin or as a web tool to run in web pages in conjunction with the \ac{klighd}~\ac{cli}.
% prev. implementation LoC: 43776
% of that, manually written LoC: 10969
% .osgiviz: 8385 Xtend
% .ls: 161 Xtend
% osgiviz.model 723 Xcore -> 21783 generated Java from that
% osgimodel.model 147 Xcore -> 11024 generated Java from that
% model.generator 1553 Java

% generated implementation LoC: 
% of that, would-have-to-be-written-manually LoC:
% .osgiviz.viz 8085 Xtend
% .ls 104 Xtend
% .osgiviz.model 256 Xcore -> 20062 generated Java from that
% .osgi.model 62 Xcore -> 5805 generated Java from that
% .osgi.generate 931 Java, of that 465 already in template

% to get that, write DSL LoC: 76
% and manual additional Java/Xtend LoC in the generator: 466

\subsubsection{Maven and Spring \acs{di}}
\label{sec:examples-maven-spring}
A second example visualization using \ac{spviz} has the goal to visualize two kinds of connections that may occur within Java projects---dependencies between \emph{modules} as defined by its Maven\footnote{
    \url{https://maven.apache.org/}
} build system,
as well as dependencies and provisions of \emph{service components} and \emph{interfaces} using \ac{di} as defined in the Spring Framework\footnote{
    \url{https://spring.io/projects/spring-framework}
}.
These connections are structured via \emph{Maven artifacts} bundling together multiple child modules.
We want to clarify to not confuse these Maven artifacts with the artifacts of our \ac{a3m}.
The modeled Maven artifacts are one kind of artifact as defined by the \ac{a3m}.
This modeled structure is similar to the \ac{osgi} example and shows that different systems can be modeled and therefore visualized.
The main difference here is how both examples have their different artifacts persisted in the file system.
For the Maven modules and artifacts, the generator parses the \emph{pom.xml} files to build the \ac{pm} of the hierarchy and the module dependencies.
The \ac{di} artifacts are extracted by parsing the Java files looking for interfaces and classes with \texttt{@Inject} and \texttt{@Named} annotations.

% Industrial partner project 3: 34/66 modules, 175 service artifacts
This example was designed by our industrial partner and verified by them with a project consisting of 34 modules and 175 service artifacts.
Their goal was to use such a visualization tool for automatically updating documentation and for onboarding of new developers and maintenance.
Other projects can also reuse parts of this design and model extractor, as the project structure based on the Maven build and the \ac{di} information are separate and can be used with other systems.
This example does not visualize all possible \ac{di} configurations of the Spring Framework, as the use case of our partner only uses the annotation-based configuration of components and requirements via \texttt{@Inject} and \texttt{@Named} annotations directly in the Java code.
Therefore this example supports to extract these annotations, but no further \ac{xml}-based configuration or other annotations.

\subsubsection{Gradle}
\label{sec:examples-gradle}
Many projects use build tools such as Gradle\footnote{
    \url{https://gradle.org/}
} to define the project structure, dependencies, and to automate the build process and tests.
This build system is extensible and features a look into the project structure, dependencies, etc. as a direct part of the build process,
thus making it easier to reuse that data for analysis tools.
Designing the \acp{dsl} and generating the visualization code for any project architecture can be done quite quickly.
However, programming an extractor for any project style can require deeper insight into the project structure and its build tools and used frameworks.
Reusing the data provided by frameworks and build tools such as Gradle can cut down the cost to develop such an extractor and also make it less error-prone.
With Gradle, data can be extracted by the same tool that is used to execute or build the project.

This Gradle example was verified with the open source project Spring Boot\footnote{
    \url{https://github.com/spring-projects/spring-boot}
}, which contains a total of 187 so-called \emph{projects} plus 729 further dependent projects.
% SpringBoot: 187/916 projects
To make this example work together with its Gradle build system, we added a task to be executed for each sub-project during the build. This collects the information of the \emph{Configuration \ac{api}} of Gradle and presents it to the model generator in a more accessible \ac{json} format.
Similarly, other code miners and extensible project build systems can be used to extract the project information for other project types.
This example shows the feasibility of using the build system to our advantage and can be combined with other hierarchies as shown in the other examples.

\subsubsection{Yarn and InversifyJS}
\label{sec:examples-yarn-inversify}
Finally, to show another full visualization combining dependencies and services outside of the Java programming language, we combined the dependency management of TypeScript projects using Yarn\footnote{
    \url{https://yarnpkg.com/}
} with the InversifyJS\footnote{
    \url{https://github.com/inversify/InversifyJS}
} \ac{di} framework.
The artifacts for this example are parsed by analyzing the \emph{yarn.lock} file generated by Yarn to extract the packages and their dependencies
and by parsing the TypeScript source files to search for interface-to-class bindings and \texttt{@inject} and \texttt{@injectable} annotations to create the hierarchy as described in the other examples.

We verified this with the open source framework Sprotty\footnote{
    \url{https://github.com/eclipse/sprotty}
} with a dependency tree of 906 packages and 126 defined service artifacts.
% Sprotty: 4/906 packages, 126 service artifacts
An example view showing the configuration of injected services as they are required by the \emph{SvgExporter} class is shown in \autoref{fig:view-sprotty-exporter}.
Here every other layer represents either a class (green), which was configured to provide its service to some interface, or an interface (blue) that is injected to a class.
Being a quite small framework, InversifyJS has no tool to visualize such hierarchies to our knowledge, so this is a good showcase to implement novel visualizations.
However, this example shows a possible usability issue.
The \ac{pm} is defined as an \ac{emf} model, which can be best extracted in Java, as the framework supports programmatically building the model, whereas writing generators in other languages would require manually conforming to the \ac{xmi} definition of the \ac{pm}, which is more effort than using the framework directly.
Future work can circumvent this issue by providing or reusing another exchange format.

\subsection{User Stories and Industry Feedback}
\label{sec:evaluation-feedback}
\ac{spviz} can be used by varying user groups with different goals and desires for a tool solution.
Next, we present and analyze three such groups and their implications for the tool design.

As a first user group we identify the \emph{software developers}, or \emph{end users}.
They want to learn and understand the system they are developing to be able to improve and extend it.
For this, they need reliable and up-to-date information about the system and a way to filter that information to some context.
Furthermore, the effort for acquiring such information should be low.
A technical solution for the software developers should allow for different representations~\cite{MalavoltaLPT+13}.
The information should furthermore update automatically, for example by integration into the build process, to lower the effort to use the solution and always have up-to-date information.
Finally, the solution should be close to or integrated into their development \ac{ide} or the documentation~\cite{ChartersTM03}
to avoid bloating their workflows.

In the second user group we identify the \emph{technology experts} as the tool designers for individual architectures.
They need a tool that is tailored to their domain technology (\eg \ac{osgi}).
If there is no such tool, or a tool is not specific enough, development of a new one should be a one-time-effort with low maintenance cost.
A solution should therefore enable experts to convert their domain knowledge into a usable tool and extract the data from the underlying project.
Its setup should furthermore be easy and understandable and work with any technology.

Lastly, we identify the \emph{software architects} as a third user group.
They need to be able to configure views and highlight parts of projects.
They also want to integrate such views in the documentation and presentations for on-boarding new developers.
Their solution requirement is that view should be interactively configurable, that this configuration can be persisted and that views based on such a configuration automatically update to changes in the underlying system.

Users can be in multiple of these groups and therefore require a combined solution.
We gathered feedback from our industrial partner on the usage of the \ac{osgi} and Maven + Spring \ac{di} examples, being applied to internal projects.
We interviewed two participants, the product owner and one of the architects of the projects, which are summarized here.
One of them fits in the software architect and partly in the software developer user group,
while the other fits in the technology expert and software architect groups.

One visualization goal the participants want to solve is to \emph{explore} the modules of their architecture to get an overview, either overall or from some specific view point.
Another goal is to \emph{explain} the architecture and specific hierarchies to others by creating architectural descriptions, without the need to update such descriptions manually.
Both participants stated that previously such diagrams were crafted and updated by hand.
While there are many visualizations out there, this shows that at least for this questionnaire the architects were not happy with other tools they used so far.
Other tools did not provide exactly what was required, because they were not usable as well, or because the architects did not find the right tool yet.

Both had the problem that views for larger projects start to require more effort to use and that clustering or pooling of artifacts into categories can induce a better hierarchical view on parts of the system.
This is especially the case when there are many artifacts of the same type being visualized in the same view.
The artifact views and category connections we described help to find the right context,
as long as the model provides enough context via some categorizations.
Further ideas indicate that diagram layouts can become a little too large for what is shown, which can be solved in future work.

Overall, their feedback indicates that the tool can and already has been used to understand parts of different system architectures.
As mentioned above and further discussed in \autoref{sec:discussion}, some improvements regarding the actual views and their interaction can be added, though that does not impair the proposed approach to create  visualizations for any project.

%%%%%%%%%%%%%%%%%%%%%%%%%%%%%%%%%%%%%%%%%%%%%%%%%%%%%%%%%%%%%%%%%%%%%%%%%%%%%%%%%%%%%%%%%%%%%%%%%%
% ---------------------------------------- Related Work ---------------------------------------- %
%%%%%%%%%%%%%%%%%%%%%%%%%%%%%%%%%%%%%%%%%%%%%%%%%%%%%%%%%%%%%%%%%%%%%%%%%%%%%%%%%%%%%%%%%%%%%%%%%%
\section{Further Related Work}
\label{sec:related}
As elaborated in the following, many related proposals go in the direction of visualization for software architectures or using meta models to describe and visualize architectures.

Following definitions used for software architecture, our tool mostly adheres to them as well and supports tooling for software architectures.
Citing Shaw and Garlan~\cite{ShawG96}, software architecture
\enquote{involves the description of elements from which systems are built,
    interactions among those elements,
    patterns that guide their composition,
    and constraints on these patterns}.
Especially the composition and interaction among any element of systems is what we also try to focus on, fitting to this definition of elements as we allow them to be as fine-granular as methods and statements.
Bass \etal~\cite{BassCK03} summarize multiple definitions, quoting that \enquote{architecture is high-level, the overall structure, structure of components and their interrelations, or components and connectors,}
indicating that our visualizations fit into the category of architecture visualizations.

According to Clements \etal~\cite{ClementsGLN+03},
\enquote{modern software architecture practice embraces the concept of architectural views. A view is a representation of a set of system elements and relations associated with them.}
Views are the main part in which architectures are perceived and mostly also designed.
We here interpret a view differently, in that it is a visual representation that emerges from the underlying architecture.
We aim to let users of \ac{spviz} use an explicit architecture following these definitions, or implicit connections through third parties through dependencies.

Shahin \etal~\cite{ShahinLK10} show how understanding architectural design decisions as communicated by software architects needs to be visualized.
They are however manually created and do not allow the extraction of data for new visualizations from the code alone.
Shahin \etal~\cite{ShahinLAB14} further discuss software architecture visualization techniques and highlight how the architectural design, architectural patterns, and architectural design decisions influence what the real code will result in and how tool support is critical for practical applicability of their visualization.

Architectures of projects are often described by \acfp{adl} in the literature.
Medvidovic and Taylor~\cite{MedvidovicT00} classify and describe the use of \acp{adl} in general.
Using their words, at one end of the spectrum, it can be argued that the primary role of architectural descriptions is to aid understanding and communication about a software system.
At the other end of the spectrum, the tendency has been to provide formal syntax and semantics of \acp{adl}, powerful analysis tools, model checkers, parsers, compilers, code synthesis tools, runtime support tools, and so on.
Our approach is not an \ac{adl}, but a way to describe project-specific architecture descriptions to create an easier step-in into creating project-specific visualizations, or a \emph{meta \ac{adl}}.
\ac{spviz} can be used for existing software architectures, \acp{adl} and \acp{mil}~\cite{PrietoDiazN86}.

A related tool for general architecture visualization using an \ac{adl} is described in work by Buchgeher \etal and Weinreich \etal~\cite{BuchgeherW08, WeinreichMBK12}.
They \enquote{emphasize the architectural description
    of a system as a central element during the whole software
    development process. The main idea is to incorporate a
    formalized architecture description model into a software
    system which is always kept up-to-date during software development.}
While they focus on continuous co-development of the software and a formal architecture description of it,
they require such a direct integration and only partly enable the analysis of legacy code that was not initially designed with their tool.
Furthermore, as their tooling is restricted to the Eclipse \ac{ide},
its use for projects developed outside the Eclipse environment is limited.
However, they allow constraint analysis to automatically detect architectural errors and inconsistencies and have immediate visual feedback on system change,
giving them some advantages in such an integrated system.
We want to bridge the gap of generic architecture \acp{adl} and an easy way to describe meaningful visualizations for them.
Our tool can interface with existing architectures to provide visualizations, while not requiring any concrete language or syntax on the architecture description itself.
Other \acp{adl} tightly integrated into specific languages, requiring the underlying projects to conform to their constraints are for example ArchJava~\cite{AldrichCN02} for Java and Codoc~\cite{BangL21} for Python.
There are other similar tools for domain-specific graphical modeling tools, partly based on \acp{dsl} applicable to different architectures.

Nimeta ~\cite{Riva04} is a tool for architecture reconstruction based on views.
They build graphical views based on so-called \emph{view-points} for arbitrary descriptions.
They clearly split the data extraction from the visualization step to allow different tools to visualize the same data,
whereas we with \ac{spviz} integrate the architectural description in the view descriptions, allowing for further filtering based on the architecture.

The CINCO tool~\cite{NaujokatLKS18} generates domain-specific graphical modeling tools from abstract specifications.
They use a similar meta meta modeling hierarchy to describe architectures.
Their use case, however, is to generate a full modeling suite for such a model with the graphs editing to modify the underlying models.
They do not aim to visualize existing architectures in a way \ac{spviz} does.
Other graphical editors for models such as Epsilon~\cite{KolvosPP06b}, Sirius~\cite{VujovicMP14}, and Spray~\cite{GerhartB16} map domain-specific models to graphical views, but require explicit descriptions on how views should be displayed and model elements should be connected in the first place.
The models to graphical views concept is similar to the KLighD framework that we use,
while we utilize the KLighD framework to make the way of connecting elements more implicit via the \ac{vc3m} \ac{dsl}.
These related papers style different parts of the visualizations for more configurability.
The \ac{spviz} approach may also support similar individual stylings beyond the current box and arrow graphs as shown above, but the tool currently does not support such individual element styling.

Another term under which visualizing architecture is understood is the reconstruction of software architecture from the area of reverse engineering.
El-Boussaidi \etal~\cite{ElBoussaidiBVM12} use the \ac{kdm}~\cite{OMG-KDM16} to describe legacy projects to visualize them,
while other use clustering algorithms to try to infer architectural meaning from otherwise non-structured code~\cite{StoermerOBV03, Riva02, Wiggerts97}.
We think approaches like these are a good way to reverse engineer and structure unstructured legacy code
which can be combinable with our visualization techniques, if they can output the results to be imported in some \ac{pm} generator described with our approach.

Petre~\cite{Petre02} analyzed what expert programmers want from visualizations.
\Eg, they need to be specific enough to get good insight into the architecture
and allow a good focus on parts that are currently interesting for the user.
Gallagher \etal~\cite{GallagherHM08} analyzed key areas for software visualization practice such as representation, navigation and interaction.
We analyzed the visualizations and interactions with the \ac{osgi} tool in our previous work~\cite{RentzDvH20}.
Most of that analysis is also applicable to the generalization of this paper.
However, the specificity and its focus are dependent on the concrete \ac{a2m} and \ac{vc2m}.
In general, we allow for the tools to be as specific as they need to be to satisfy the specificity requirement.

%%%%%%%%%%%%%%%%%%%%%%%%%%%%%%%%%%%%%%%%%%%%%%%%%%%%%%%%%%%%%%%%%%%%%%%%%%%%%%%%%%%%%%%%%%%%%%%%%%
% ----------------------------------------- Discussion ----------------------------------------- %
%%%%%%%%%%%%%%%%%%%%%%%%%%%%%%%%%%%%%%%%%%%%%%%%%%%%%%%%%%%%%%%%%%%%%%%%%%%%%%%%%%%%%%%%%%%%%%%%%%
\section{Discussion}
\label{sec:discussion}
While this paper presents our proposed concept for describing arbitrary architectures and their visualization,
some parts in the configurability, the visualization, and the handling of the \acp{dsl} can be improved in the future.
We would like the generated visualization modules to be extensible to make the final visualizations more configurable to fit aesthetics criteria of the architecture to visualize.
A first idea is to open up an \ac{api} that allows the individual renderings for the overviews, the artifacts, or even the connections or the arrow heads to be configurable.
This could be achieved for example via some injection mechanism similar to the configurability of many functional aspects in the Xtext framework or an extension to the \ac{dsl} syntax.
This could also enable the modification of automatic artifact colors.
Another part that could be configurable is mapping the visualized artifacts to some web-link leading to documentation of that artifact or being integrated into different \acp{ide} such as Eclipse and \ac{vscode}.

An improvement to the tooling to design a new visualization would be to allow using the \acp{dsl} outside of Eclipse in \eg \ac{vscode} or a command line tool, and generate the required code from there.
Furthermore, support to write a model generator using other languages could help developers.
Another improvement would be to make the \ac{a3m} compatible with other meta models such as the \ac{kdm}~\cite{OMG-KDM16}.

In the \ac{vc3m} \ac{dsl} we would like to enable further configuration.
The bundle dependencies view in \autoref{lst:vc3m-dsl-example} only permits a direct connection of bundles to other bundles.
However, bundles may be indirectly connected via a different \emph{package} artifact, whose connections should be shown, without showing the package artifacts themselves.
An extension to the \ac{dsl} could allow writing something like \lstinline[style=spviz]|connect OSGi.Bundle via OSGi.Bundle.PackageDependency|, to indicate that the parent bundle artifact of the connected package dependency should be connected instead.
A different syntax to better distinguish it from the category connection could be used as well.
Another extension would enable (recursive) connections of filtered artifacts in artifact views.
The bundle dependencies artifact view of products in \autoref{lst:vc3m-dsl-example} for example could show the project-known dependencies of all filtered bundles with some syntax such as \lstinline[style=spviz]|OSGi.Bundle from OSGi.Product>OSGi.Bundle and connected|.
This would not only show the bundle hierarchy of all bundles defined in the product, but also their (recursive) dependencies, as they are always part of a deployed product anyways.

To solve another issue that the texts for the artifact names can get quite long and bloat the required size for the diagram, we would like to add an option to only show hints of the names and expand the names with hover feedback or to use automatic label management to reduce the layout footprint of the individual artifacts~\cite{SchulzeLvH16}.

Lastly, the stored \ac{vcm} could be improved as well.
Currently a stored \ac{vcm} restores the last visualized state by the \acp{id} of each artifact and their connections to make it possible that a change in the project updates the model and restores the previous view.
For a \ac{vcm} which was configured to show all connections of some specific element,
while an update to the project introduces some new connected element, that connection will not be shown in the updated visualization.
Instead, the port which was white before to indicate that there are no further elements to connect to would be black in the updated version, indicating there are further connections currently not shown.
This change is not easily recognizable.
However, restoring the \ac{vcm} by connecting all elements where all elements were connected before could mitigate that issue.

To address threats to validity of the industry feedback,
that part of the evaluation is not meant to be the final study to validate the usability of our proposed SPViz tool. The questionnaire was not structured in a controlled manner and is meant to be viewed as a initial argument towards showing the usefulness of \ac{spviz} for generating customized visualization tools.
We plan on conducting a further user study investigating how users design and adapt new visualization tools using our proposed approach in the future.

%%%%%%%%%%%%%%%%%%%%%%%%%%%%%%%%%%%%%%%%%%%%%%%%%%%%%%%%%%%%%%%%%%%%%%%%%%%%%%%%%%%%%%%%%%%%%%%%%%
% ----------------------------------------- Conclusion ----------------------------------------- %
%%%%%%%%%%%%%%%%%%%%%%%%%%%%%%%%%%%%%%%%%%%%%%%%%%%%%%%%%%%%%%%%%%%%%%%%%%%%%%%%%%%%%%%%%%%%%%%%%%
\section{Conclusion}
\label{sec:conclusion}
\ac{spviz} is a new approach for software architects to quickly create a visualization tool
they can use to explore any otherwise obscure architecture.
It enables its users to create automatically updating architectural views for documentation purposes and to explain relations to others.
We built a tool following this approach to make a previously presented approach for visualizing, exploring, and documenting \ac{osgi} projects available for arbitrary software architectures, highlighting the usability of such a concept.
The visualizations use state-of-the-art and well-accepted views on connections within software systems such as dependencies and service structures.
We compared the tool to other meta modeling tools and architectural visualizations, such as \acp{adl}, which usually require projects to adapt to.
We do not require projects to use any specific architecture, but support the description of the architecture for any project.
\ac{spviz} can be used as a visualization tool generator for legacy systems to visualize specific parts that other tools do not cover.
It can also be used to quickly set up a visualization for new and emerging languages and system structures.
To be applicable to projects that have no real own architecture and are just a collection of source files,
a combination with other tools clustering and organizing specific artifacts is recommended.

Overall, the tool has been used and evaluated on multiple projects, showing its benefits.
However, as discussed in \autoref{sec:discussion}, some areas can still be improved in future research to widen the use cases of this architecture-agnostic software visualization tool generator.

\bibliographystyle{IEEEtran}
\bibliography{../bib/cau-rt,../bib/pub-rts,../bib/rts-arbeiten, blinded-refs}

\end{document}